%% file: simplifying-note.tex
\documentclass[11pt, headinclude=true]{scrartcl}

\usepackage{lmodern}
\usepackage{babel} 

\usepackage{graphicx}
\usepackage{enumerate}

\usepackage{amssymb}
\usepackage{amsmath}
\usepackage{amsthm}
\usepackage{dsfont}
\usepackage{bm}
\usepackage{todonotes}
\usepackage[round, comma]{natbib}

\usepackage{subfig}

\usepackage{hyperref}
\hypersetup{
	colorlinks   = true,
	urlcolor     = LMUpurple,
	linkcolor    = LMUgreen,
	citecolor   = LMUgreen
}
\usepackage[capitalize,nameinlink]{cleveref}

\usepackage{tikz}
\usetikzlibrary{shapes,arrows,decorations.pathmorphing,graphs,positioning,backgrounds}

\input{latex-templates/layout.tex}
\input{latex-templates/lmu-colors.tex}

\input{latex-templates/microtype.tex}
\input{latex-templates/theorems-en.tex}

\input{latex-shortcuts/math.tex}

\input{latex-shortcuts/stats.tex}
\input{latex-shortcuts/tex.tex}

\usepackage{lipsum}  

\title{Simplified vine copula models: state of science and affairs}
\author{Thomas Nagler\thanks{Department of Statistics, LMU Munich, Akademiestr.~7, 80799 Munich, Germany, \texttt{mail@tnagler.com}}}

\begin{document}

\maketitle

\begin{abstract}
	Vine copula models have become highly popular practical tools for modeling multivariate dependencies. To maintain tractability, a commonly employed simplifying assumption is that conditional copulas remain unchanged by the conditioning variables. This assumption has sparked a somewhat polarizing debate within the copula community. The fact that much of this dispute occurs outside the public record has placed the field in an unfortunate position, impeding scientific progress. In this article, I will review what we know about the flexibility and limitations of simplified vine copula models, explore the broader implications, and offer my own, hopefully reconciling, perspective on the issue. \\[11pt]
\end{abstract}


\section{Introduction}

Copula models provide a general framework for capturing dependencies between variables and have been widely adopted across various disciplines, including finance, insurance, environmental science, and medicine. Vine copulas---also known as pair-copula constructions---have gained particular prominence. These models break down high-dimensional dependence structures into a sequence of bivariate (conditional) copulas. Today, vine copulas are applied extensively, with hundreds of publications emerging each year.
At the core of most works on vine copulas is the simplifying assumption: conditional dependencies in the model do not vary with the conditioning variables. This assumption improves both statistical and computational tractability but costs some flexibility.

The simplifying assumption has become an oddly polarizing issue within our community, although the tension is not reflected in the published scientific record. Papers highlighting its limitations are dismissed in peer review for relying on  `unrealistic toy examples.' On the other side, methodological works on simplified models face rejections with the recommendation to abandon research involving the simplifying assumption altogether because it might be the `next formula that kills Wall Street', referring to a famous article on the Gaussian copula \citep{salmon2012formula}. To my knowledge, neither of these positions has been voiced, let alone justified in the public scientific record. This state of affairs is quite unfortunate, as it hinders scientific progress.

The main purpose of this note is to bring the debate into the open scientific discourse and to offer my own perspective on the matter. At the very least, it should provide a good basis for a more constructive discourse going forward. The key points are as follows: 
\begin{itemize} 
	\item Simplified vine copulas are widely used and continue to drive meaningful scientific advancements across various fields. Therefore, research that enhances understanding, capability, or applicability of these models remains valuable. 
	\item Like all models, simplified vine copulas are approximations of reality. In high-stakes applications, we must exercise caution with inferences and critically evaluate our assumptions. As such, we should welcome any theoretical or empirical work that sheds light on the limitations of simplified models. 
\end{itemize}

To support these assertions, I will review what we know about simplified models, place the assumption into a broader context, and discuss some of its implications. This note is intentionally kept non-technical, skipping over certain details to maintain focus on the broader perspective.

\section{Background}

\subsection{Copula models}

Suppose we are interested in modeling the distribution of a random vector $\bm{X} \in \mathbb{R}^d$. For simplicity, we will assume throughout that $\bm{X}$ is absolutely continuous. Let $F_1, \dots, F_d$ denote the marginal distributions of $X_1, \dots, X_d$, and let $F$ denote their joint distribution. Sklar's theorem states that there exists a unique function $C\colon [0,1]^d \to [0,1]$, called a copula, such that
\begin{align*}
	F(x_1, \ldots, x_d) = C(F_1(x_1), \ldots, F_d(x_d)).
\end{align*}
In fact, $C$ is the joint distribution of the random vector $\bm{U} = (F_1(X_1), \ldots, F_d(X_d))$, which has uniform marginals. The marginals $F_j$ describe the individual behavior of $X_j$, while the copula $C$ captures their dependence. Sklar's theorem can also be expressed in terms of densities:
\begin{align*}
	f(x_1, \ldots, x_d) = c(F_1(x_1), \ldots, F_d(x_d)) \times \prod_{j = 1}^d f_j(x_j),
\end{align*}
where $f$, $c$, and $f_j$ are the densities corresponding to $F$, $C$, and $F_j$, respectively.

\subsection{The simplifying assumption in the bivariate case}

Now consider the following conditional version of Sklar's theorem \citep{patton2001modelling}. For any index set $D \subset \{1, \ldots, d\}$, indices $i, j \notin D$, and every value $\bm x_D \in \R^{|D|}$, the conditional distribution of $(X_i, X_j)$ given $\bm X_D = \bm x_D$ can be expressed as
\begin{align*}
	F_{i, j| D}(x_i, x_j \mid \bm x_D) = C_{i, j|D}(F_{i| D}(x_i \mid \bm x_D), F_{j| D}(x_j \mid \bm x_D) \mid \bm x_D),
\end{align*}
where $C_{i, j|D}(\cdot \mid \bm x_D)$ is a bivariate copula for every value of $\bx_D$.\footnote{In the vine copula literature, the conditional copula is more commonly denoted by $C_{i,j;D}$ to distinguish it from the conditional distribution of $(U_i, U_j)$ given $\bm U_D$, which is not a copula. The latter distribution or related conditionals do not appear in this article, so there should be no confusion.} More precisely, it is the conditional joint distribution function of the random vector $(U_{i|D}, U_{j|D})$ given $\bm X_D = \bm x_D$, where 
$$U_{i|D} = F_{i|D}(X_i \mid \bm X_D), \quad U_{j|D} = F_{j|D}(X_j \mid \bm X_D).$$
The operation $U_{i|D} = F_{i|D}(X_i \mid \bm X_D)$ takes $X_i$ and discards all information that $\bm X_D$ provides for its marginal distribution. To see this, compute
\begin{align*}
	\Pr(F_{i|D}(X_i \mid \bm X_D) \le u \mid  \bm X_D = \bm x_D) 
	= \Pr(X_i \le F^{-1}_{i|D}(u \mid \bm x_D) \mid  \bm X_D = \bm x_D) &= u,
\end{align*}
so $U_{i|D} \mid \bX_{D} = \bm x_{D} \sim \mathrm{Unif}[0, 1]$, and we obtain $U_{j|D}  \mid \bX_{D} = \bm x_{D}  \sim \mathrm{Unif}[0, 1] $ by the same argument. This also implies $U_{j|D} \indep \bm X_D$ and $U_{j|D} \indep \bX_D$, but not that $(U_{i|D}, U_{j|D}) \indep \bm X_D$.
The copula $C_{i,j|D}(\cdot \mid \bm x_D)$ then captures the dependence between $X_i$ and $X_j$, after all marginal associations with $\bm X_D$ have been removed. 
If no dependence remains after this step, the variables are conditionally independent given $\bm X_D$ and $C_{i, j|D}(\cdot \mid \bm x_D)$ is the independence copula, irrespective of the value of $\bm x_D$. If some dependence \emph{does} remain, the conditional copula $C(\cdot \mid \bm x_D)$ may change with $\bm x_D$. From a modeling perspective, this poses a big challenge, especially if $|D|$ is large.
To alleviate this, one may make the following assumption:
\begin{quote}
	\textbf{Simplifying assumption}: The conditional copula $C_{i, j|D}(\cdot \mid \bm x_D)$ does not depend on $\bm x_D$.
\end{quote}

The Gaussian distribution provides an accessible analogy. Let $\bm X \sim \Ncal(0, \Sigma)$ and define $\eps_{i|D}$ and $\eps_{j|D}$ as the erorrs from regressing $X_i$ and $X_j$ on $\bm X_D$,  scaled to unit variance. This is, in essence, what the operation $F_{i|D}(X_i \mid \bm X_D)$ does. The conditional copula $C_{i, j|D}(\cdot \mid \bm x_D)$ characterizes the dependence of the regression erorrs $\eps_{i|D}$ and $\eps_{j|D}$. In the Gaussian world, dependence is synonymous with correlation, so we can think of $C_{i, j|D}(\cdot \mid \bm x_D)$ as the \emph{conditional correlation coefficient} $\rho_{i, j|D}(\bm x_D) = \corr(\eps_{i|D}, \eps_{j|D} \mid \bm X_D = \bm x_D)$. The simplifying assumption asserts that this correlation coefficient does not change with the value $\bm x_D$.  Indeed, the Gaussian distribution satisfies $\rho_{i, j|D} \equiv \rho_{i, j;D} $, where $ \rho_{i, j;D}$ is the \emph{partial correlation coefficient} of $(X_i, X_j)$ given $\bm X_D$. This equality does not typically hold outside of Gaussian models, however. For more details, we refer to  \citet{gijbels2015partial} and \citet{ spanhel2016partial} who expand on this idea in terms of conditional and partial copulas.

\subsection{The simplifying assumption in vine copulas}

Modern vine copula models are built on the foundation set by the seminal works of \citet{joe1996, cooke1997markov, bedford2001probability, bedford2002vines}, and \citet{aas2009pair}. For a more comprehensive introduction, we refer to the introductory article by \citet{czado2022vine} or the textbook by \citet{czado2019analyzing}.
Vine models break down the joint distribution of $\bm X$ into its marginal distributions and several conditional bivariate copulas. The conditioning must adhere to specific rules, which are summarized in a graphical structure known as a \emph{regular vine}.

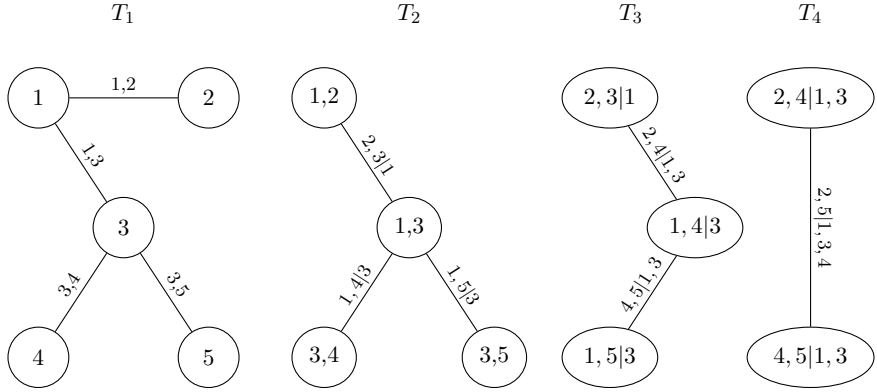
\begin{figure}
	\centering 
		\tikzstyle{VineNode} = [ellipse, fill = white, draw = black, text = black, align = center, minimum height = 1cm, minimum width = 1cm, scale = 0.8]
		\tikzstyle{DummyNode}  = [draw = none, fill = none, text = white] 
		\tikzstyle{TreeLabels} = [draw = none, fill = none, text = black] 
		\newcommand{\labelsize}{\footnotesize} 
		\newcommand{\yshiftLabel}{-0.3cm}  
		\newcommand{\yshiftNodes}{-0.75cm} 
		\newcommand{\xshiftTree}{0.5cm}    
		\newcommand{\rotateLabels}{-57}    
			\centering
			\begin{tikzpicture}	[every node/.style = VineNode, node distance =1.4cm] 
			\node (1){1}
			node[DummyNode]  (Dummy12)   [right of = 1]{} 
			node             (2)         [right of = Dummy12] {2}
			node             (3)         [below of = Dummy12, yshift = \yshiftNodes] {3}
			node[DummyNode]  (Dummy45)   [below of = 3, yshift = \yshiftNodes]{}
			node             (4)         [left of = Dummy45] {4}
				node             (5)         [right of = Dummy45] {5}
				node             (12)        [right of = 2, xshift = \xshiftTree] {1,2} %
				node[DummyNode]  (Dummy12x)  [right of = 12]{} 
			node             (13)        [below of = Dummy12x, yshift = \yshiftNodes] {1,3}
			node[DummyNode]  (Dummy45c3) [below of = 13, yshift = \yshiftNodes]{} 
			node             (34)        [left  of = Dummy45c3] {3,4}
				node             (35)        [right of = Dummy45c3] {3,5}
				node             (15c3)      [right of = 35, xshift = \xshiftTree] {$1,5|3$}
			node[DummyNode]  (Dummy15c3x)[right of = 15c3]{}
			node             (14c3)      [above of = Dummy15c3x, yshift = -\yshiftNodes] {$1,4|3$}
			node[DummyNode]  (Dummy23c1x)[above of = 14c3, yshift = -\yshiftNodes]{}
			node             (23c1)      [left of  = Dummy23c1x] {$2,3|1$}	
			node             (24c13)     [right of = Dummy23c1x, xshift = \xshiftTree] {$2,4|1,3$}
				node             (45c13)     [right of = Dummy15c3x, xshift = \xshiftTree] {$4,5|1,3$}
			node[TreeLabels] (T1)        [above of = Dummy12] {$T_1$}
			node[TreeLabels] (T2)        [above of = Dummy12x] {$T_2$}
			node[TreeLabels] (T3)        [above of = 23c1] {\hspace{0.7cm}$T_3$} 
			node[TreeLabels] (T4)        [above of = 24c13] {$T_4$}	 
			;	    	
			\draw (1) to node[draw=none, fill = none, font = \labelsize,
													above, yshift = \yshiftLabel] {1,2} (2);
			\draw (1) to node[draw=none, fill = none, font = \labelsize, 
													rotate = \rotateLabels, above, yshift = \yshiftLabel] {1,3} (3);   
			\draw (3) to node[draw=none, fill = none, font = \labelsize, 
													rotate = \rotateLabels, above, yshift = \yshiftLabel] {3,5} (5);  
			\draw (3) to node[draw=none, fill = none, font = \labelsize, 
													rotate = -\rotateLabels, above, yshift = \yshiftLabel] {3,4} (4); 
			\draw (12) to node[draw=none, fill = none, font = \labelsize, above, 
													rotate = \rotateLabels, above, yshift = \yshiftLabel] {$2,3|1$} (13);   
			\draw (13) to node[draw=none, fill = none, font = \labelsize, above, 
													rotate = \rotateLabels, above, yshift = \yshiftLabel] {$1,5|3$} (35); 
			\draw (13) to node[draw=none, fill = none, font = \labelsize, above, 
													rotate = -\rotateLabels, above, yshift = \yshiftLabel] {$1,4|3$} (34); 
			\draw (23c1) to node[draw=none, fill = none, font = \labelsize, above, 
													rotate = \rotateLabels, above, yshift = \yshiftLabel] {$2,4|1,3$} (14c3);   
			\draw (14c3) to node[draw=none, fill = none, font = \labelsize, above, 
													rotate = -\rotateLabels, above, yshift = \yshiftLabel] {$4,5|1,3$} (15c3); 
			\draw (24c13) to node[draw=none, fill = none, font = \labelsize, above, 
													rotate = -90, above, yshift = \yshiftLabel] {$2,5|1,3,4$} (45c13);
			\end{tikzpicture}
	\caption{A regular vine structure with $d = 5$ variables.}
	\label{fig:vine-tree}
\end{figure}

In simple terms, a vine is a sequence of trees $\Vcal = (T_1, \ldots, T_{d-1})$, each comprising a set of nodes $V_k$ and edges $E_k$. Each edge in the sequence is associated with a conditional pair of variables $a_e, b_e$ and a conditioning set $D_e$. The conditioning set $D_e$ for $e \in T_k$ has $k - 1$ elements. As we move along the trees, pairs of variables are conditioned on an increasing number of other variables.
An exemplary vine graph is shown in \cref{fig:vine-tree}. 
We have the following result \citep[Theorem 3]{bedford2001probability}.

\begin{theorem} For any joint density $f$ and regular vine $\Vcal = (T_1, \ldots, T_{d-1})$,
	\begin{align}\label{eq:rvine_density}
	\begin{aligned}
		&\quad \; f(x_1,\ldots x_d) \\
		& =   \prod_{j = 1}^d f_j(x_j) 	\times  \prod_{k = 1}^{d-1} \prod_{e \in E_k} c_{a_e, b_e| D_e} (F_{a_e \vert D_e}(x_{a_e}\vert \bx_{D_e}), F_{b_e\vert D_e}(x_{b_e} \vert \bx_{D_e})\ \mid \bm x_{D_e}).
	\end{aligned}
	\end{align}
\end{theorem}

The theorem states that the joint density of $\bm X$ can be expressed as the product of the marginals and a cascade of conditional bivariate copulas. The copulas $c_{a_e, b_e| D_e}$ are the copula densities associated with conditional distribution of $(X_{a_e}, X_{b_e})$ given $\bm X_{D_e}$. Equivalently, $c_{a_e, b_e| D_e}$ is the copula density of the random vector $(U_{a_e \vert D_e}, U_{b_e \vert D_e}) = (F_{a_e \vert D_e}(X_{a_e} \vert \bm X_{D_e}), F_{b_e \vert D_e}(X_{b_e} \vert \bm X_{D_e}))$ given $\bm X_{D_e}$.

\begin{figure}
	\centering 
	\fig[0.9]{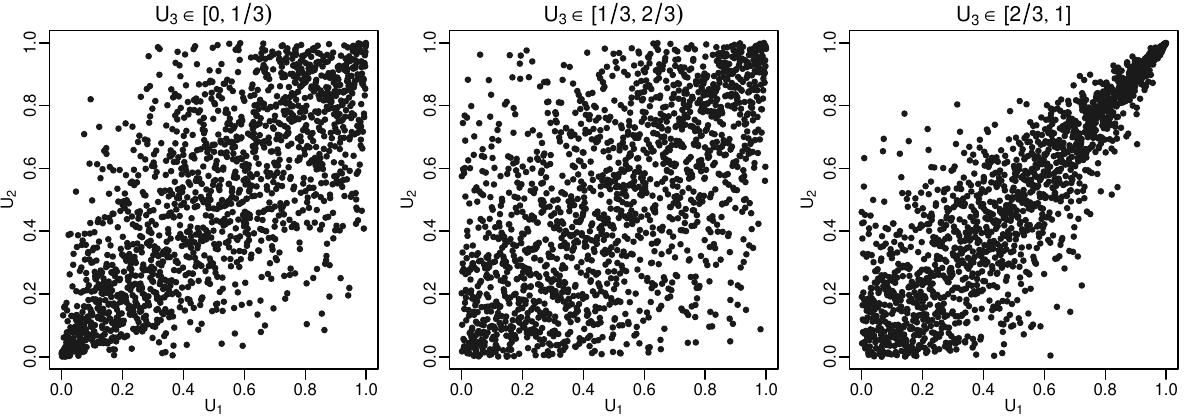}
	\caption{Whether the simplifying assumption holds depends on the vine structure. The scatter plots show a violation of the assumption in samples from the conditional copula $c_{1, 2|3}$ when restricting $U_3$ to different intervals. Data were simulated from a simplified vine copula $c_{1, 2} \times c_{2, 3} \times c_{1, 3| 2}$, where all pair-copulas are Gumbel with Kendall's $\tau = 0.8$.}
	\label{fig:structure-vs-sa}
\end{figure} 

The decomposition \eqref{eq:rvine_density} turns into a model as soon as we restrict the set of marginal densities and conditional copulas in a way that makes the formula statistically and computationally tractable.
One such restriction is the simplifying assumption: the conditional copulas do not depend on the conditioning value $\bm x_{D_e}$. Importantly, the assumption only restricts conditional copulas that are explicit in the model, implicit ones may still deviate from it.
So to be precise, the simplifying assumption in vine copula models reads as follows.
\begin{quote}
	\textbf{Simplifying assumption in vine copulas}: A distribution $F$ satisfies the simplifying assumption with respect to the vine structure $\Vcal$, if \eqref{eq:rvine_density} holds with
	\begin{align*}
		c_{a_e, b_e|D_e}(\cdot \mid \bx_{D_e}) = c_{a_e, b_e |D_e}(\cdot) \quad \text{for all } \bx_{D_e} \in \R^d \mbox{ and } e \in \Vcal.
	\end{align*}
\end{quote}
This hopefully also clarifies a common misconception. Simplified vines do not assert that all conditional dependencies are constant. They only assert that the few dependencies that are explicit in the tree structure are constant. And curiously: except for special cases, decomposing a $\Vcal$-simplified model along a different vine structure $\Vcal'$ leads to violations of the simplifying assumption. \cref{fig:structure-vs-sa} provides an illustration, where data is simulated from a model satisfying the simplifying assumption for the decomposition $c_{1, 2} \times c_{2, 3} \times c_{1, 3| 2}$, but not for the decomposition $c_{1, 3} \times c_{2, 3} \times c_{1, 2| 3}$.

Spare few exceptions, practical implementations of simplified vine copula models further constrain marginals and/or copulas to a fixed selection of convenient parametric families. Despite the many restrictions, such models are remarkably flexible considering other parametric alternatives. Each marginal distribution can be specified individually, whether light or heavy tailed, symmetric or skewed, unimodal or multimodal, unbounded or bounded support. Similarly, each (explicit) partial dependence can be chosen individually to reflect different dependence strengths, tail behavior, or symmetries. At the same time, simplified vines remain computationally tractable, even with hundreds of variables.

\section{What we know about the limitations of simplified models}

So what exactly do we know about the flexibility and limitations of simplified vine copula models? In what follows, I recall what I consider the most important insights from the literature and provide some of my own interpretation.

\subsection{The simplifying assumption in common dependence models}

\subsubsection{Known simplified models}

A natural questions is which known multivariate models satisfy the simplifying assumption and has been investigated by \citet{stoeber2013simplified}. The authors show that the simplifying assumption is satisfied by the multivariate Gaussian and Student's $t$ distributions. These are the basis of some of the most commonly used statistical models, so 
the simplifying assumption does not seem entirely unreasonable. However, these two are the only scale-mixtures of Gaussians that satisfy it. Additionally, the multivariate Clayton copula is the only Archimedean copula that satisfies the assumption.
This list is very short. While the simplifying assumption may appear natural in the context of vine copulas, it is quite unusual considering other common ways of generating multivariate models. Some caution is therefore warranted. 


\subsubsection{Archimedean dependence}

\begin{figure}
	\centering
	\includegraphics[width=0.4\textwidth]{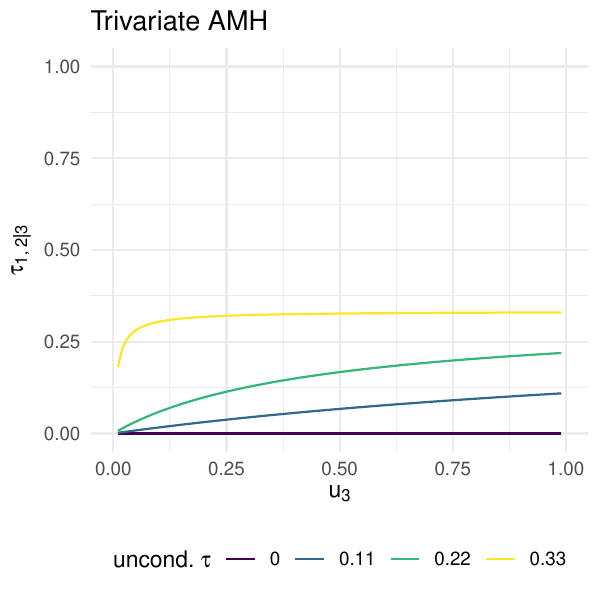} \hspace{0.66cm}
	\includegraphics[width=0.4\textwidth]{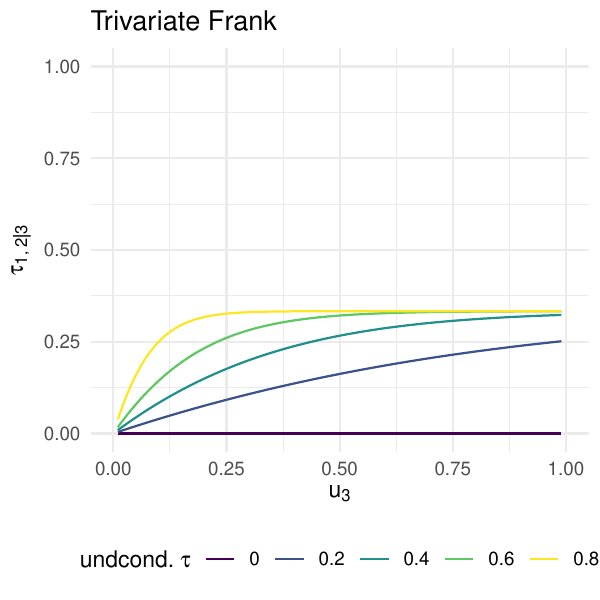}
	\caption{Conditional Kendall's tau when the trivariate distribution is either of AMH or Frank type with given unconditional Kendall's $\tau$.}
	\label{fig:tau-conditional}
\end{figure}

\citet{MESFIOUI2008372} investigated conditional copulas associated with Archimedean models in more detail. For example, they show that a trivariate Frank copula $c$ with parameter $\eta$ has conditional $c_{1, 2| 3}(\cdot \mid u_3)$ of the Ali-Mikhail-Haq (AMH) type with parameter $1 - \exp(- \eta u_3) $. Another example is that the conditional $c_{1, 2| 3}(\cdot \mid u_3)$ of a trivariate AMH copula with parameter $\eta$ is again of AMH type with parameter $u_3 \eta / (1 - \eta + u_3 \eta)$.
\cref{fig:tau-conditional} shows the conditional Kendall's $\tau$ associated with the copula $ c_{1, 2| 3}(\cdot \mid u_3)$, when the trivariate copula $c$ is given by either the AMH or Frank family with a given conditional Kendall's $\tau$. We observe that the conditional Kendall's $\tau$ does indeed change with the value of the conditional variable $U_3$ (except for the limiting case of trivariate independence), so the simplifying assumption is violated. However, the deviations from constant $\tau$ are not too severe, not least because the conditional Kendall's $\tau$ is bounded to $[0, 1/3]$ for all Archimedean copulas \citep{MESFIOUI2008372}.

It's not difficult to think of situations where a simplified vine is a much more sensible model than the Archimedean alternatives. Examples are settings where $\tau_{1, 2 | 3} \gg 1/3$ or the dependence is not exchangeable. Whether the simplifying assumption holds is not the most urging question here. We should rather ask whether a simplified vine provides a better approximation of the true dependence structure than the one implied by the Archimedean copula.

\subsubsection{Extremal dependence}

A second observation in \cref{fig:tau-conditional} is that the conditional $\tau$ tends to flatten out as it becomes large. This relates to another set of recent results investigating extremal dependence.
\citet{kiriliouk2023x} explore vine decompositions of \emph{tail copulas}. Tail copulas arise in extreme value theory as characterization of dependence of extreme events. The \emph{lower tail copula} $R \colon \R_{> 0}^d \to \R$ of a copula $C$ is defined as 
\begin{align*}
	R(\bx) = \lim_{t \to 0} \frac{C(tx_1, \dots, tx_d)}{t}.
\end{align*}
Despite the name, the lower tail copula is not a copula, but a function that characterizes the dependence in the lower tail of the distribution. 
The lower tail copula is best known from the lower tail dependence coefficient
\begin{align*}
	\lambda_L = \lim_{t \to 0} \frac{C(t, t)}{t} = R(1, 1).
\end{align*}
The density corresponding to the tail copula is the \emph{tail copula density} \citep{LI201399} 
$$r(\bx) = \lim_{t \to 0}\frac{c(tx_1, \dots, tx_d)}{t^{d - 1}},$$
where $c$ is the density associated with $C$. 
\citet{kiriliouk2023x} proved that tail copula densities can be decomposed along a vine tree as 
\begin{align*}
	r(\bx) = \prod_{e \in E_1}r_{a_e, b_e}(x_{a_e}, x_{b_e}) \times \prod_{k = 2}^{d-1} \prod_{e \in E_k} c_{a_e, b_e| D_e}^*(R_{a_e | D_e}(x_{a_e} | \bx_{D_e}), R_{b_e | D_e}(x_{b_e} | \bx_{D_e}) \mid \bx_{D_e}),
\end{align*}
where $r_{a_e, b_e}$ are bivariate tail copula densities, $c_{a_e, b_e| D_e}^*$ conditional copula densities, and $R_{i|D}$ conditional distribution functions associated with $R$. We use an asterisk to indicate that $c_{a_e, b_e| D_e}^*$ is not the same as $c_{a_e, b_e| D_e}$ in \eqref{eq:rvine_density}, because they are associated with different random variables.

In general, the conditional copula densities $c_{a_e, b_e| D_e}^*(\cdot \mid \bx_{D_e})$ may depend on the conditioning value $\bx_{D_e}$. However, \citet{kiriliouk2023x} show that the simplifying assumption holds for some the most widely used  models for extremal dependence. One example is the Hüsler-Reiss model \citep{husler1989maxima}, which arises as the extremal limit of a Gaussian copula model with correlation $\rho \to 1$ at an appropriate rate. This may not be too surprising, because the Gaussian copula satisfies the simplifying assumption. A second class satisfying the simplifying assumption is the \emph{scaled extremal Dirichlet model}. This is a broad family including many widely used special cases, such as the negative logistic model of \citet{JOE199075}, the logistic model of \citet{emile1960distributions}, and the extremal Dirichlet model by \citet{coles1991modelling}.

It appears that the simplifying assumption may sometimes be justified in the context of extremal dependence models. This should not be misinterpreted as a general endorsement of simplified vine copulas for estimating extreme event probabilities. Tail copulas are objects different from copulas and have to be fitted only on the most extreme subset of observations. A simplified vine model fit to all observations can still be off, because the estimated parameters may underrepresent the dependence strength in tail.

\subsection{Empirical evidence for and against the simplifying assumption}

Another route of investigation is to check for deviations from the simplifying assumption empirically on real data. Several tests and estimation procedures for simplified models have been developed in the literature. Many works include brief illustrations with real data sets\footnote{Examples where time is a covariate are excluded deliberately, since we would not treat it as a random variable in multivariate probabilistic models.}:
\begin{itemize}
	\item \citet{acar2013statistical} show that in data from twins, the dependence between their birth weights change with the gestational age. On a second data set, the dependence between blood pressure measurements at distinct times is not affected by the change in body mass index between periods.
	\item \citet{gijbels2017nonparametric} and \citet{gijbels2021omnibus} show that the assumption is violated for a data set of characteristics of different types of concrete.
	\item \citet{acar2012beyond} and \citet{GIJBELS2017111} show that the dependence between concentrations of some chemicals in water measurements change with the concentration of some of the others.
	\item \citet{gijbels2017nonparametric} find no violation of the simplifying assumption in the dependence of systolic and diastolic blood pressure when cholesterol is taken as a covariate.
	\item \citet{DERUMIGNY2020104610} show that the dependence between male and female life expectancy changes with the GDP of the country. 
	\item \citet{derumigny2023testing} find that the simplifying assumption can not be rejected for filtered returns of the Eurostoxx50 and S\&P500 indices conditioned on the previous day, but is rejected when considering events spanning multiple days. This should be taken with a grain of salt, as the effects of pre-filtering and potentially left-over serial dependence are ignored.
\end{itemize}
The aforementioned works develop methodology for testing or estimating covariate effects in conditional copulas. It is unsurprising that they find violations of the simplifying assumption, since the data sets are chosen to illustrate the methodology. But this is a  warning shot. The simplifying assumption is far from automatic, and sometimes violated severely.

Recall that in a concrete vine copula model, the simplifying assumption must only hold for the conditional pairs that the vine structure makes explicit.
Whether violations are problematic in practice depends very much on which structure we choose.
The most common way to select the structure is the heuristic of \citet{dissmann2013selecting}, which aims at capturing the strongest dependencies in the first few trees. In the example from \cref{fig:structure-vs-sa}, it would have recovered the `true' structure under which the simplifying assumption holds. It does not explicitly account for the simplifying assumption, however. At least there is hope that, because there is not much dependence left to capture, many of the higher-order conditional copulas become less relevant, making violations less problematic.

\citet{kurz2022testing} developed a test for jointly testing all explicit conditional dependencies in a vine copula model. Their empirical study investigates the validity of the assumption for structures selected by Dissman's algorithm. The test confirms a violation in the chemical data set already investigated by \citet{acar2012beyond} and \citet{GIJBELS2017111}. Apparently, the structure selected by Dissman's algorithm does not get us around the issue. The test found no violation in three further data sets comprising financial time series. This is sensible: the multivariate $t$ copula is known to be a decent model for such data, and it satisfies the simplifying assumption.
\citet{kraus2017growing} modified Dissman's algorithm to account for $p$-values of tests for the simplifying assumption. They show that some violations of the assumption in the chemical data set from \citet{acar2012beyond} and \citet{GIJBELS2017111} can be circumvented by an appropriate choice of structure, but the algorithm cannot resolve all violations.

The aforementioned works provide anecdotal evidence that whether and how severely the simplifying assumption is violated in practice depends on the data at hand. From a philosophical point of view, there isn't much reason to believe it is \emph{exactly} true---apart from the special case of conditional independence. Any consistent test will detect the smallest deviation if the sample size is large enough, and the above references all use at least several hundred observations both in real-data analyses and power studies. 

The same can be said about any parametric model for the dependence and related goodness-of-fit tests. That does not mean that simplified or parametric models are inherently bad. What is important is that we understand the limitations of our models and the consequences of their violations.

\subsection{Issues with interpreting simplified models}

\citet{spanhel2019simplified} raised a potential issue when it comes to interpreting simplified models. To set things up, we have to distinguish more carefully between the theoretical construct of a simplified vine copula and its practical implementation. Most often, simplified vine models are fit to the data in a step-wise fashion. First, select and estimate the pair-copulas in the first tree by maximum likelihood. Given these estimates, select and estimate the pair-copulas in the second tree, and so on.
To reflect this procedure in theoretical considerations, \citet{spanhel2019simplified} introduced the \emph{partial vine copula (PVC)}. It is constructed as a simplified vine copula, where the pair-copulas are chosen to sequentially minimize the Kullback-Leibler divergence, given the choices in previous trees. If the simplifying assumption holds, the PVC coincides with the true copula. If the assumption does not hold, the PVC is generally different from the best approximation (in the KL sense) among all simplified vine copulas. This distinction is important, because we usually estimate the PVC in practice. 

The difference between the PVC and the best approximation highlights two concerning issues.
First, the best simplified vine approximation may not contain the true pair-copulas, even for the unconditional pairs in the first tree level. When fitting a simplified model with global maximum likelihood, we must be careful with interpreting the estimated pair-copulas.
They may not reflect the true dependence structure of a given pair. The PVC on the other hand, always holds the true pair-copulas in the first tree. In the second tree however, violations of the simplifying assumption may lead to a misinterpretation of the estimated dependence. In general, we may over- or underestimate the strength of the dependence in certain regions. The copula associated with each pair only gives us a sense of on-average dependence. Another issue appears from the third tree on. The PVC may assign conditionally independent variables a pair-copula far from independence. Here, the PVC merely corrects for non-simplified patterns we have missed out on in the previous trees.

\begin{figure}
	\centering 
	\fig[0.4]{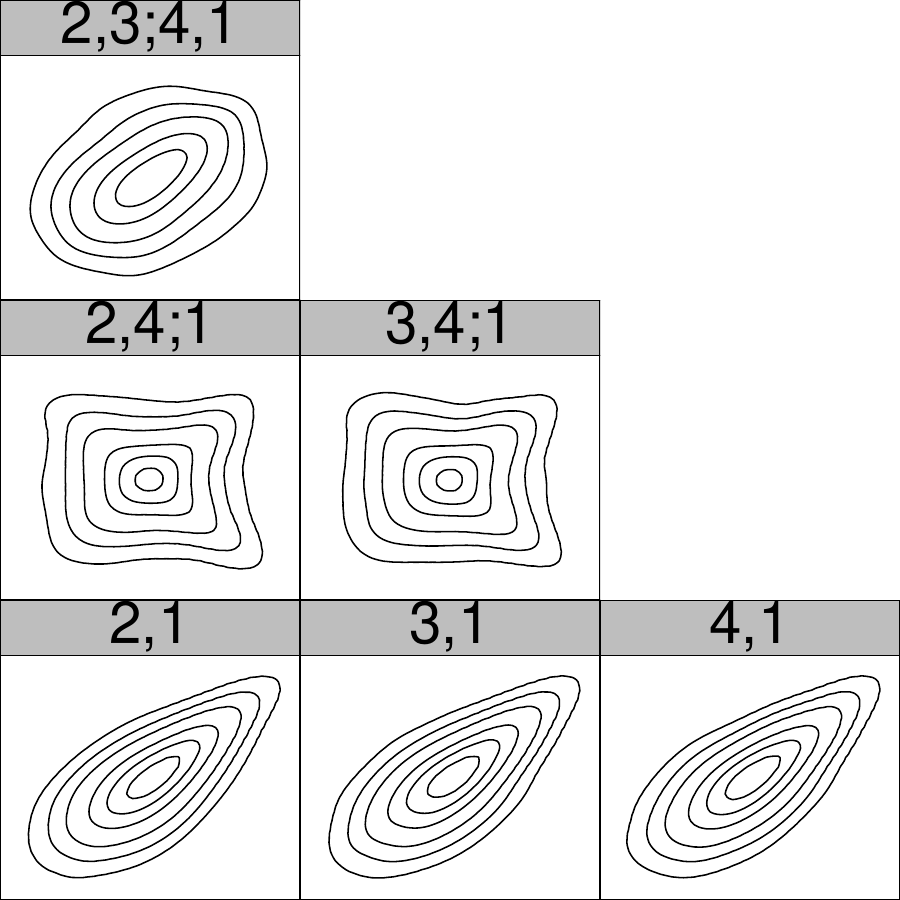}  \hspace*{0.66cm} \fig[0.4]{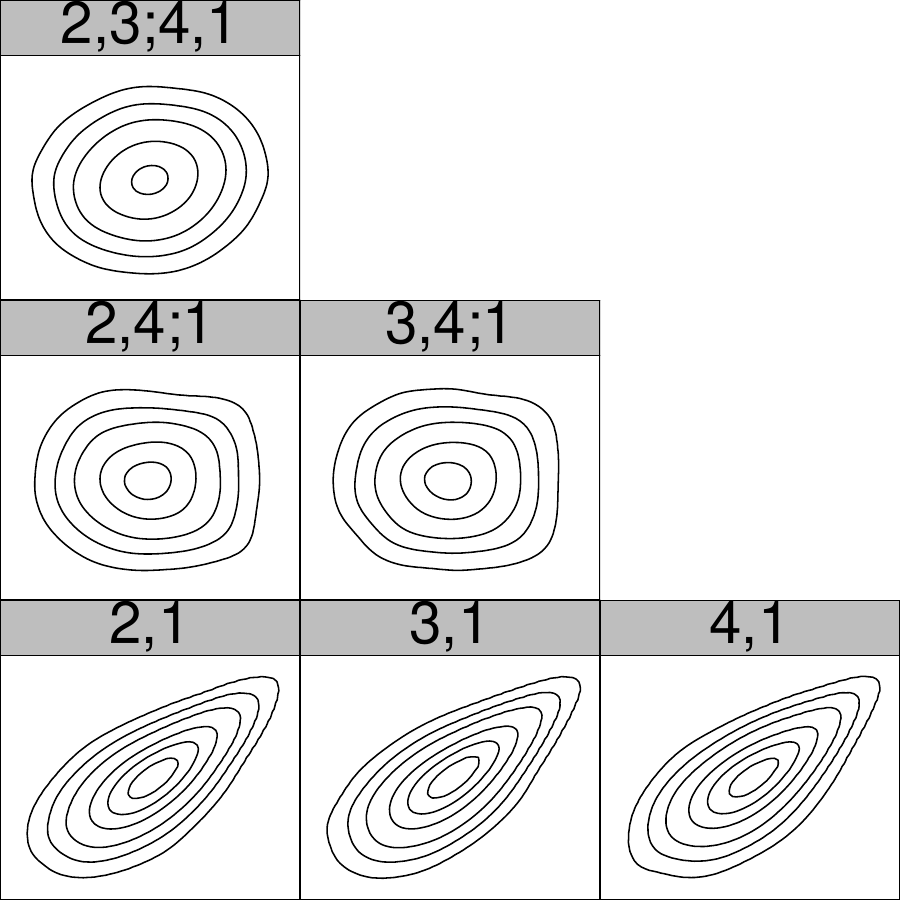} 
	\caption{Contour plots of estimated pair-copulas in a simplified vine fit to non-simplified data (left: strong violation, right: moderate violation). While $X_2$ and $X_3$ are conditionally independent given $X_1, X_4$, the partial vine copula induces spurious positive dependence (panels in the top row).}
	\label{fig:contour-PVC}
\end{figure} 

To illustrate this, consider the following example:
\begin{itemize}
	\item Specify the copulas in the first tree with $c_{1, 2} = c_{1, 3} = c_{1, 4}$ as a Gumbel copula with Kendall's $\tau = 0.5$;
	\item Specify conditional copulas in the second tree with $c_{2, 4|1}(\cdot \mid u) = c_{3, 4|1}(\cdot \mid u)$ as a Gumbel copula with Kendall's $\tau$ equal to $0.9 (2 u - 1)$ (strong violation) or $0.4 (2u - 1)$ (moderate violation), rotating the copula by 90 degrees if $\tau < 0$.
	\item The copula in the last tree, $c_{3, 2 | 1, 4}$, we leave at independence. 
\end{itemize}
We now simulate $10 \, 000$ observations from the model and fit a nonparametric simplified vine copula using the \texttt{rvinecopulib} package \citep{rvinecopulib}.
The fitted model on the left shows strong positive dependence in the pair $(X_2, X_3)$ given $(X_1, X_4)$ (top row), even though they are conditionally independent. Also for the moderate violation (right plot), there is weak positive correlation with a Kendall's $\tau$ of approximately 0.08, which passes usual statistical significance thresholds. Concluding that the variables are conditionally dependent would clearly be a mistake. For the moderate violation, the pair-copulas in the second tree further seem close to independence, because they average out positive and negative dependencies. Concluding that the variables are conditionally independent is also a mistake.

This leads us to the following conclusion. It is generally good advice to refrain from interpreting fitted conditional dependencies, as long as we haven't found evidence that the simplifying assumption holds.

\subsection{The approximation error of simplified models}

Another relevant question is how big the approximation error of simplified vine copula models can be. The approximation error is the difference between the true copula and the copula implied by the model. This error may be measured in several distance metrics that have different interpretations and may lead to different conclusions. 

The approximation error was analyzed formally in a seminal paper by \citet{mroz2021}. Their most surprising result is that the collection of simplified vine copulas is dense in the set of all copulas with respect to the uniform metric $d_\infty(C, D) = \sup_{\bu \in [0, 1]^d} |C(\bu) - D(\bu)|$. This means that the approximation error can be made arbitrarily small by choosing the right simplified model. We must not misinterpret this result as saying that the simplifying assumption is inconsequential in practice. 
The proof of the result is based on the fact that all empirical copulas can be represented as simplified vine copulas, with independence pair-copulas everywhere but for the first tree. The same argument would suggest that Markov tree distributions \citep[e.g.,][]{ansari2024comparisonresultsmarkovtree} are dense in the set of all distributions. However, assuming universal conditional independence is clearly restrictive in a practical sense.  And indeed, this all breaks down when more statistically meaningful metrics are considered: \citet{mroz2021} show that simplified vine copulas are nowhere dense with respect to the topology induced by weak convergence of conditional distributions, the total variation metric, and the Kullback-Leibler divergence. 

The aforementioned results concern the distance of any simplified vine copula to the true copula. As discussed in the previous section, simplified vine copulas are typically constructed in a step-wise fashion. \citet{mroz2021} go further and show that the partial vine copula (PVC) can have $d_\infty$-distance to the true copula of more than 1/8. In three dimensions, this corresponds to more than 28\% of the diameter of the space. This is substantial, so a simplified model may lead to a very poor approximation of the true copula if the violation is extreme. 

The theoretical arguments of \citet{spanhel2019simplified} and \citet{mroz2021} are based on ground truth models for the copula that are unlikely to be encountered in real data. 
For example, the $d_\infty$ distance between the true copula and the fitted PVC in the simulation model from the previous section can be estimated\footnote{This number is based on simulating $10\,000$ samples from the true and fitted model and comparing empirical probabilities.} to around 0.06 for the strong violation and 0.03 for the moderate violation. Although the deviations from the simplifying assumption are substantial, these numbers are much smaller than the 1/8 bound computed in \citet{mroz2021}. 
It would be a huge mistake to dismiss the practical implications of their results based on this, however. It is good practice to look for theoretical examples that are both simple and maximally bad for the model. This allows to understand the deviation's effects and avoids cluttering mathematical arguments with unnecessary technicality. This lends the results an instructive value beyond the bare computation of numbers. And even if the $d_\infty$-distance is `just' 0.03, underestimating the probability of an event by 0.03 can be catastrophical if the event is rare but consequential. One should carefully assess the validity and implications of the simplifying assumption when stakes are high.

It would be an equally big mistake to dismiss all models making the simplifying assumption based on these findings.
The results are missing important context about alternative models. For example, I am not aware of similar results providing the approximation error of (hierarchical) Archimedean models, a popular alternative to vine copulas. Even though they do not make the simplifying assumption, we should expect this error to be much larger. As mentioned earlier, the conditional dependence in such models changes in very specific ways (with conditional Kendall's $\tau$ bounded by 1/3), and it wouldn't be difficult to construct examples that deviate in  bad ways (for example, vine copulas containing conditional copulas with Kendall's $\tau$ much larger than $1/3$ or much less than $0$). This comes on top of inherent symmetry properties of such models that are often violated in practice. For example, consider a hierarchical Archimedean model fit using the \texttt{HAC} package \citep{HAC} with default settings to the simulated data from the previous section. While the $d_\infty$ distances from the true model are similar to the simplified vine, their KL divergence is substantially large:$\approx 0.89$ (HAC) vs.~$\approx 0.67$ (vine) for the strong violation; $\approx 0.28$ (HAC) vs.~$\approx 0.18$ (vine) for the moderate violation. The purpose of this example is not to argue that vine copulas are better than Archimedean models, but to highlight that any model has its limitations and that the simplifying assumption is not the only one to be concerned about.

When there is evidence for violations of the simplifying assumption, methods for estimating non-simplified vine copula models \citep{ vatter2018generalized, schellhase2018estimating} may provide a potential alternative, at least in moderate dimension. Such models are substantially less tractable, however. Further research is needed to adequately assess the limitations of comparable alternatives, and potentially find better ones.

\subsection{Simplified vines as statistical models}

Simplified vine copulas are statistical models for the dependence structure of a random vector. From a philosophical point of view, they are no different from other statistical models serving similar purposes. Models are approximations of reality, and as such, mostly wrong. To paraphrase a famous quote by George Box: The question is not whether they are wrong, but whether they are useful.

Generally speaking, a statistical model is a collection $\Mcal = \{P_\theta\colon \theta \in \Theta\}$ of probability distributions $P_\theta$ characterized by a parameter $\theta$ ranging over some set $\Theta$. If $\Theta$ is finite-dimensional, we call the model parametric, otherwise nonparametric. The model is used to make inferences about the true distribution $P_0$ of the data. 
In practice, statistical models always serve a purpose. We may want to estimate the probabilities of some events, we may want to predict future observations, or we may want to understand specific aspects of the mechanism generating the data. 
A statistical model is useful if it serves its purpose. 
That's typically the case if:
\begin{enumerate}[(i)]
	\item there is $\theta_0 \in \Theta$ such that $P_{\theta_0}$ provides a good approximation of $P_0$ --- for the aspects of the distribution we care about;
	\item the model is computationally tractable and allows for efficient estimation of the parameter $\theta_0$.
\end{enumerate}

Point (i) highlights that we are rarely interested in the global approximation quality of $P_{\theta_0}$. If the goal is to estimate lower tail probabilities, we should not worry much about the approximation quality in the upper tail. If the goal is to predict $Y$ given covariates $\bX$, we don't care much about the marginal distributions of $\bX$. If the goal is to derive simple explanations for the relationship between $Y$ and $\bX$, a fully nonparametric model may be too complex to be useful. 
This urges statisticians to carefully consider what aspects of the model are important and which simplifications are justified to achieve their goal.
A simple example where a deviation from the simplifying assumption may not be problematic is given in the following.
\begin{example}
	To give a simple example, where a deviation from the simplifying assumption is not problematic, consider the following regression model: 
	\begin{align*}
		 Y = X_1 + X_2\eps, \quad \text{where} \quad (X_1, X_2, \eps) \sim \Ncal(0, I_3).
	\end{align*}
	The conditional distribution of $(Y, X_1)$ given $X_2 = x_2$ is a bivariate Gaussian with correlation $\rho(x_2) = 1 / (1 + x_2^2)^{1/2}$, so the simplifying assumption is violated. However, the regression function is $\E[Y \mid X_1 = x_1, X_2 = x_2] = x_1$, which is the same as in the simplified model $(Y, X_1, X_2) \sim \Ncal(0, \Sigma)$ with 
	\begin{align*}
		\Sigma = \begin{pmatrix}
			2 & 1 & 0 \\
			1 & 1 & 0 \\
			0 & 0 & 1
		\end{pmatrix}.
	\end{align*}
	In fact, both the marginal family for $Y$ and the copula are misspecified here. The situation changes if we are interested in conditional quantiles, where ignoring the simplifying assumption may lead to substantial errors.
\end{example}

Point (ii) emphasizes that we cannot disentangle the approximation qualities of a model from our ability to estimate it. Generally speaking, more flexible models are harder to estimate---the ubiquitous bias-variance trade-off. 
Specifically, suppose we are interested in some aspect $T$ of the true distribution $P_0$, such as extreme event probabilities or conditional quantiles. We can decompose the total error as follows:
\begin{align*}
	T(P_{\hat \theta_n}) - T(P_0) = \underbrace{T(P_{\wh \theta_n}) - T(P_{\theta_0})}_{\text{estimation error}} + \underbrace{T(P_{\theta_0}) - T(P_0)}_{\text{approximation error} },
\end{align*}
where $\wh \theta_n$ is the estimated parameter based on $n$ observations. The estimation error usually decreases as the sample size $n$ grows, but the approximation error depends only on the model and the true distribution. Larger sample sizes thus allow us to use more flexible models, such as nonparametric or non-simplified ones.

In extreme cases, the model can be so large that consistent estimation is impossible. This is the case, for example, if $\Theta$ is the collection of all probability densities on $\R^d$. Parametric families of densities on the other hand can be estimated at $O_p(n^{-1/2})$ rate under mild conditions. Merely assuming that the densities are twice continuously differentiable allows for estimation at $O_p(n^{-2/(d + 4)})$ rate. 
This rate becomes extremely slow when $d$ is even moderately large. \citet{nagler2016evading} showed that the simplifying assumption alleviates this. In particular, they construct nonparametric estimators of the partial vine copula that converge at $O_p(n^{-1/3})$ rate --- irrespective of the dimension $d$. The simplifying assumption effectively makes estimation as easy as in a bivariate model. This comes at the cost of a potentially large approximation error, but it may be a good trade-off in practice. \citet{nagler2016evading} show in simulations that, even on large samples, an estimated simplified model can be much closer to the true density than a fully nonparametric estimate, even if the simplifying assumption is violated severely. The gains in estimability may simply outweigh the loss in approximation ability of the simplified model.

\section{Conclusion}

The simplifying assumption is a powerful tool for achieving computational and statistical efficiency while maintaining a high degree of flexibility. Of course, flexibility is to be understood in relative terms, compared to alternative models granting similar tractability. Simplified vines have proven immensely useful in numerous scientific fields. 
In many application areas, simplified vine copulas are among the most sophisticated and best fitting models known and accessible to field experts. Despite the simplifying assumption not always being true, these models are clearly a huge improvement over, for example, a multivariate Gaussian model (which also satisfies the simplifying assumption). While this may sound like a low bar for people from the copula community, Gaussianity often forms the foundation of state-of-the-art methods in the application domain. 

With part of the same reasoning, similar praise could be sung for generalized linear regression models. They are extremely useful in many fields but have their limitations. We know just about everything about these limitations. Which deviations cause which effect; how bad can a linear approximation be; how to test and correct for these deviations. The same cannot be said about the simplifying assumption. Practitioners tend to not worry about such an abstract assumption in an already complicated model. Copula experts may have a good intuition about what is happening, but intuition is not the same as scientific knowledge, which is currently scarce.  A non-exhaustive list of important questions that remain to be answered is:
\begin{itemize}
	\item How bad are violations of the simplifying assumption if focusing on parametric models?
	\item How do we appropriately measure/estimate the approximation error of a specific simplified vine copula model to the true copula?
	\item How can we mitigate the effects of violations of the simplifying assumption in a computationally and statistically efficient way?
	\item Which downstream tasks are insensitive to violations of the simplifying assumption, which are not?
\end{itemize}
Research that fills some of this gap is important and should be cherished.

Given our current knowledge about the limitations of the simplifying assumption, it seems appropriate to end with some recommendations for practitioners working with simplified vine copulas:
\begin{itemize}
	\item \emph{Reflect} on what goal the model serves, what aspects are most important, and how detrimental violations of the simplifying assumption would be.
	
	\item \emph{Evaluate and compare} to other state-of-the-art models to see how the behavior is different and whether the simplified vine model improves in the aspects that matter for the application.
	
	\item \emph{Test} for the simplifying assumption if its validity is relevant. This is certainly the case when stakes are high or the interpretation of pair-copulas in higher trees is important. Several R packages are available for testing the assumption, visual diagnostics, or non-simplified vine models; see \citet{pacotest, LocalCop, gamCopula, CondCopulas}.
\end{itemize}

\bibliographystyle{apalike}
\bibliography{ref}

\end{document}

%% file: latex-templates/layout.tex
\usepackage{setspace}
\usepackage[
  left=1.3in,
  right=1.3in,
  top=0.7in,
  bottom=0.7in,
  includeheadfoot
]{geometry}

%% file: latex-templates/lmu-colors.tex
\usepackage{xcolor}

\definecolor{LMUblue}{RGB}{0, 17, 88}
\definecolor{LMUlightblue}{RGB}{92,177,235}
\definecolor{LMUgreen}{RGB}{0,136,58}
\definecolor{LMUlightgreen}{RGB}{170, 173, 0}
\definecolor{LMUred}{RGB}{190,25,8}
\definecolor{LMUpurple}{RGB}{176, 32, 121}
\definecolor{LMUorange}{RGB}{241, 135, 0}
\definecolor{bggray}{RGB}{245, 245, 245}
\definecolor{LMUgray}{RGB}{0.2,0.2,0.2}


%% file: latex-templates/microtype.tex
\usepackage[tracking=true,expansion=true,stretch=15,shrink=15]{microtype}
\DeclareMicrotypeSet*[tracking]{my}
{ font = */*/*/sc/* }
\SetTracking{ encoding = *, shape = sc }{ 45 }
\SetProtrusion{encoding={*},family={bch},series={*},size={6,7}}
{1={ ,750},2={ ,500},3={ ,500},4={ ,500},5={ ,500},
  6={ ,500},7={ ,600},8={ ,500},9={ ,500},0={ ,500}}
\SetExtraKerning[unit=space]
{encoding={*}, family={qhv}, series={b}, size={large,Large}}
{1={-200,-200},
  \textendash={400,400}}

%% file: latex-templates/theorems-en.tex
\usepackage{shadethm}
\usepackage{etoolbox}

\newtheoremstyle{new}
  {12pt}      
  {12pt}      
  {\itshape}  
  {}          
  {\bfseries\color{black}} 
  {.}         
  { }         
  {}          
\theoremstyle{new}
\newtheorem{theorem}{Theorem}[section]

\newtheorem{example}[theorem]{Example}

\AfterEndEnvironment{theorem}{\noindent\ignorespaces}
\AfterEndEnvironment{corollary}{\noindent\ignorespaces}
\AfterEndEnvironment{proposition}{\noindent\ignorespaces}
\AfterEndEnvironment{lemma}{\noindent\ignorespaces}
\AfterEndEnvironment{definition}{\noindent\ignorespaces}
\AfterEndEnvironment{assumption}{\noindent\ignorespaces}
\AfterEndEnvironment{example}{\noindent\ignorespaces}
\AfterEndEnvironment{remark}{\noindent\ignorespaces}
\AfterEndEnvironment{claim}{\noindent\ignorespaces}

\Crefname{theorem}{Theorem}{Theorems}
\Crefname{corollary}{Corollary}{Corollaries}
\Crefname{proposition}{Proposition}{Propositions}
\Crefname{lemma}{Lemma}{Lemmas}
\Crefname{definition}{Definition}{Definitions}
\Crefname{example}{Example}{Examples}
\Crefname{remark}{Remark}{Remarks}
\Crefname{claim}{Claim}{Claims}

\usepackage[breakable, theorems, skins]{tcolorbox}
\tcbset{enhanced}

\definecolor{shadethmcolor}{cmyk}{0,0,0,0.075}    
\definecolor{shaderulecolor}{rgb}{1,1,1}   

\newtheoremstyle{shad}
  {12pt}      
  {12pt}      
  {\itshape }  
  {}          
  {\bfseries\color{black}} 
  {.}         
  { }         
  {}          
\theoremstyle{shad} 

\newshadetheorem{thmbox}[theorem]{Theorem}
\newshadetheorem{corbox}[theorem]{Corollary}
\newshadetheorem{propbox}[theorem]{Proposition}
\newshadetheorem{lembox}[theorem]{Lemma}
\newshadetheorem{exbox}[theorem]{Example}
\newshadetheorem{defbox}[theorem]{Definition}
\newshadetheorem{rembox}[theorem]{Remark}

\newtheoremstyle{shad*}
  {12pt}      
  {12pt}      
  {\itshape }  
  {}          
  {\bfseries\color{black}} 
  {.}         
  { }         
  {}          
\theoremstyle{shad*} 
\newshadetheorem{anobox}{}

\Crefname{theorem}{Theorem}{Theorems}
\Crefname{corbox}{Korollar}{Corollaries}
\Crefname{propbox}{Proposition}{Propositions}
\Crefname{lembox}{Lemma}{Lemmas}
\Crefname{exbox}{Beispiel}{Examples}
\Crefname{defbox}{Definition}{Definitions}
\Crefname{rembox}{Anmerkung}{Remarks}

%% file: latex-shortcuts/math.tex
\usepackage{bm}
\usepackage{dsfont}

\newcommand{\R}{\mathds{R}}

\newcommand{\E}{\mathds{E}}

\newcommand{\bu}{\bm{u}}

\newcommand{\bx}{\bm{x}}

\newcommand{\bX}{\bm{X}}

\newcommand{\Mcal}{\mathcal{M}}
\newcommand{\Ncal}{\mathcal{N}}

\newcommand{\Vcal}{\mathcal{V}}

\newcommand{\eps}{\varepsilon}



%% file: latex-shortcuts/stats.tex
\providecommand{\Pr}{}
\renewcommand{\Pr}{\mathbb{P}}

\newcommand{\corr}{{\mathrm{Corr}}}

\newcommand{\wh}[1]{\widehat{#1}}

\newcommand{\indep}{\perp \!\!\! \perp}

%% file: latex-shortcuts/tex.tex
\newcommand{\fig}[1][1]{
  \includegraphics[width = #1\textwidth]
}

\makeatletter
\def\blfootnote{\gdef\@thefnmark{}\@footnotetext}
\makeatother